\documentclass[prb,preprint]{revtex4}
\usepackage{graphicx}
\makeatletter
\usepackage{epsfig}
\def\@dotsep{4.5}
\usepackage{dcolumn}
\usepackage{amsmath}
\makeatother

\usepackage{graphicx}
\usepackage{bm}
\usepackage{amsfonts}
\usepackage{amsbsy}
\usepackage{amssymb}
\usepackage[mathscr]{eucal}

\newcommand{\beq}{\begin{equation}}
\newcommand{\eeq}{\end{equation}}
\newcommand{\ba}{\begin{array}}
\newcommand{\ea}{\end{array}}
\newcommand{\bea}{\begin{eqnarray}}
\newcommand{\eea}{\end{eqnarray}}
\newcommand{\bseq}{\begin{subequations}}
\newcommand{\eseq}{\end{subequations}}

\begin{document}

\title{Electron-Phonon Coupling Constant of Metallic Overlayers from Specular 
He-Atom Scattering}

\author{{ G. Benedek}$^{a,b}$}

\author{  { Salvador Miret-Art{\'e}s}$^{a,c}$}

\author{J. P. Toennies$^{d}$}

\author{J. R. Manson$^{a,e}$}

\affiliation{~\\
$^a$ Donostia International Physics Center (DIPC), Paseo Manuel de Lardiz{{a}}bal, 4,
20018 Donostia-San Sebastian, Spain
\\
$^b$ Dipartimento di Scienza dei Materiali, Universit{\`a} di Milano-Bicocca, Via Cozzi 53, 20125 Milano, Italy
\\
$^c$ Instituto de F\'isica Fundamental, Consejo Superior de Investigaciones Cient\'ificas, Serrano 123, 28006 Madrid, Spain
\\
$^{d}$ Max Planck Instit\"ut f\"ur Dynamik und Selbstorganisation, Am Fassberg 17,
37077 G\"ottingen, Germany
\\
$^{e}$ Department of Physics and Astronomy, Clemson University, Clemson, South Carolina 29634, U.S.A.
}

\date{\today}

\begin{abstract}
He atom scattering has been shown to be a sensitive probe of electron-phonon interaction properties 
at surfaces.  Here it is shown that measurements of the thermal attenuation of the specular He atom diffraction peak (the Debye-Waller effect) can determine the electron-phonon coupling constant 
$\lambda$ for ultrathin films of metal overlayers on various close-packed metal substrates.  Values 
of $\lambda$ obtained for single and multiple monolayers of alkali metals, and for Pb layers 
on Cu(111), extrapolated to large thicknesses, agree favorably with known bulk values.  
This demonstrates that He atom scattering can measure the electron-phonon coupling strength as a 
function of film thickness on a layer-by-layer basis.
\end{abstract}

\maketitle

\newpage

\vspace{2cm}
\begin{figure}
\includegraphics[width=5.5cm]{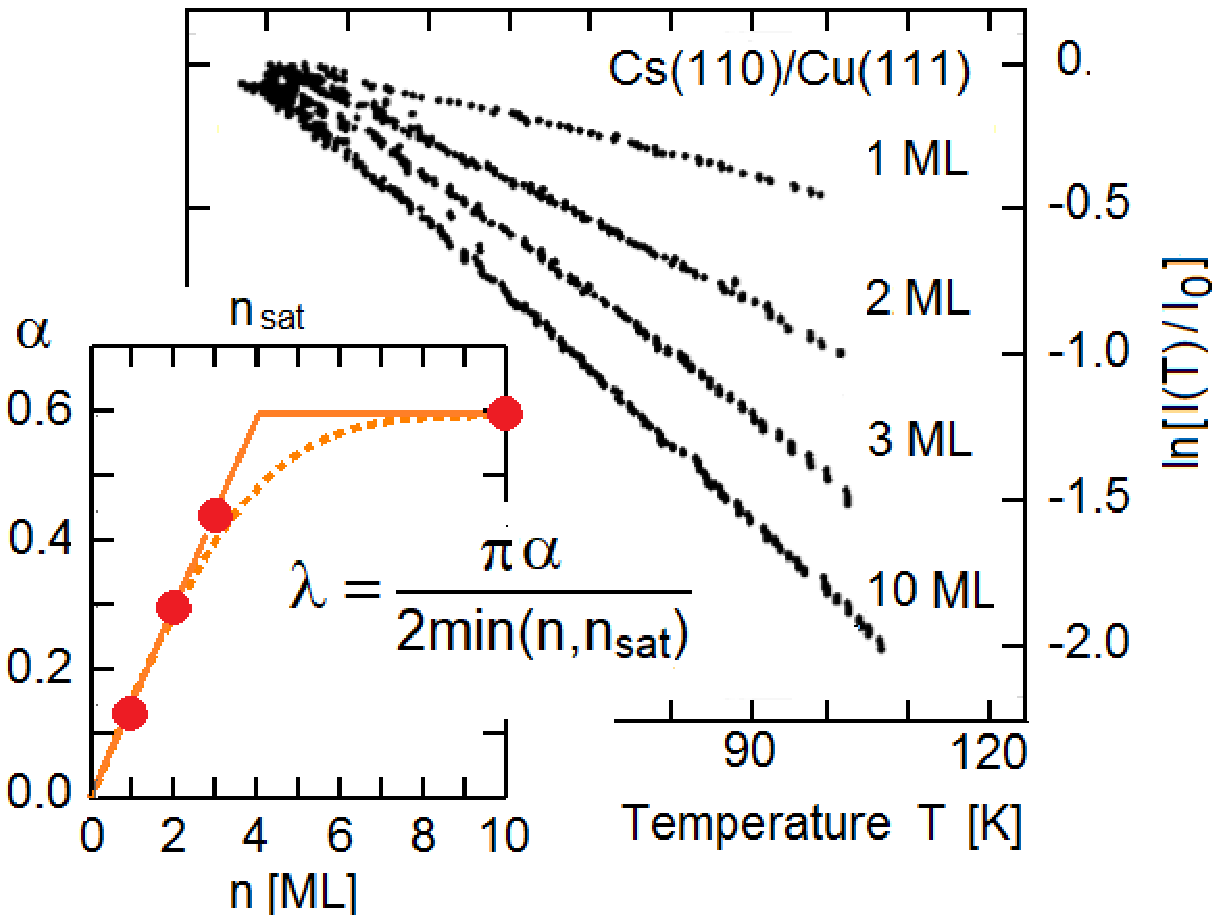}
\end{figure}





Atom scattering at low energies has a long history as a useful probe of surface properties.~\cite{Reviews2,Reviews,Farias}
The most common atomic projectiles are He atoms formed into a monoenergetic beam directed toward the surface  with energies in the thermal range, meaning kinetic energies less than 100 meV.
From ordered surfaces the scattered spectra consist of sharp diffraction peaks and a background which is composed of both elastically and inelastically reflected particles.
In such experiments measurements of elastic diffraction in the scattered spectra give information on  surface structure and order.  The
elastic part of the
diffuse intensity scattered continuously over all outgoing angles provides information on disorder such as steps, defects and adsorbates on the surface.\cite{Disorder}  Energy-resolved measurements of the inelastic spectra reveal surface phonon modes\cite{Doak} as well as localized modes of adsorbates and adsorbate layers.\cite{Bertino}  Among other features that have been measured are the cross sections of adsorbates and defects,\cite{CrossSections} as well as layer-by-layer crystal growth.\cite{Comsa}

A unique feature of low energy He atom scattering (HAS) is that the atoms scatter, not from the atomic cores of the target surface, but from the rarefied density of electron states whose wave functions extend outward in front of the outermost layer of surface atoms.  Thus, the scattered atoms sense the structure and phonon vibrations of the surface atomic cores only through their contributions to the electron density outside (roughly about 3 \AA) above the terminal surface layer.  For this reason atom-surface scattering has been demonstrated to be a sensitive probe of the electron-phonon interaction, in metals \cite{Skl} as well as in semimetals with electron and hole pockets
at the Fermi level. \cite{13Tam,13Tame,13Kra,Benedek-14} 
Inelastic He atom scattering has been shown to be proportional to the mode-selected components of the electron-phonon coupling constant $\lambda$.  For  metal crystal surfaces, inelastic scattering can even detect surface modes for which the largest core displacement amplitudes are located several layers beneath the outermost terminal layer, a property that has been called the quantum sonar effect. \cite{Skl}  The Debye-Waller factor, which describes the thermal attenuation of all quantum scattering features (diffraction, diffuse elastic intensity, and inelastic intensities) has been shown to depend in a straightforward manner on the electron-phonon coupling constant $\lambda$.\cite{Manson-JPCL-16,Manson-SurfSciRep} 
Recently, Tamtoegl et al. \cite{17Tam} have proved with state-of-the-art 3He-spin echo experiments that the method reported in Ref. \cite{Manson-JPCL-16} can be extended to the important case of surface quantum-well and Dirac states in topological insulators.

In this Letter we examine  experimental studies, using He atoms as probes, of layer-by-layer growth of ordered metallic monolayers on close-packed metal substrates.  Such experiments typically monitor the intensity of the specular beam, or a different diffraction peak, under conditions of continuous deposition of the metal of interest, and as each additional layer is completed the intensity reaches a maximum, with intervening minima occurring approximately at half-layer coverage.  We demonstrate here that these oscillations in intensity, as well as measurements of the Debye-Waller factor on completed layers, can be used to extract values of $\lambda$ associated with individual layer thicknesses.  This means that the evaluation of $\lambda$ as a function of  the thickness of the metallic overlayer can be measured by He atom scattering on a layer-by-layer basis.  Such knowledge of the electron-phonon interaction near a surface is important for understanding adsorption, chemisorption and chemical reactions at surfaces, and is certainly important in predicting surface superconductivity.

The Debye-Waller (DW) factor associated with a diffraction peak of intensity $I(T)$ is expressed as $I(T) = I_0 \exp[-2W({\bf k}_f, {\bf k}_i;T) ]$ where $I_0$ is its intensity for a frozen surface, $2W({\bf k}_f, {\bf k}_i;T)$ is the DW exponent, $T$ is the surface temperature, and ${\bf k}_f$ and ${\bf k}_i$ are the final and initial He atom wavevectors.  In a harmonically vibrating crystal at sufficiently large temperature $2W$ is linearly proportional to $T$.  At temperatures well below the Debye temperature the DW factor saturates to a value less than one and independent of $T$ as a consequence of zero point motion.  Thus, $I_0$ is the extrapolation of the intensity in the linear region of $2W$  back to $T=0$.

It has recently been shown that the electron-phonon coupling strength (or mass enhancement factor) $\lambda$ in the case of conducting surfaces can be obtained from a temperature dependent measure of the DW factor of the specular diffraction peak, in the region where $2W$ is linear, according to the relation \cite{Manson-JPCL-16,Manson-SurfSciRep}
\begin{eqnarray} \label{Eq1}
 2W({\bf k}_f, {\bf k}_i;T) ~\cong~ 4 \mathcal{N}(E_F) ~ \frac{m E_{iz}}{m_e^* \phi} ~ \lambda ~ k_B T
~,
\end {eqnarray}
where $\phi$ is the work function, $m_e^*$  is the effective electron mass, $m$ is the mass of the He atom, $k_B$ is the Boltzmann constant, and $\mathcal{N}(E_F)$ is the effective electron density of states (DOS) at the Fermi level referred to the surface unit cell and includes only those states which cause a surface charge density oscillation as a consequence of the thermal vibrational motion of the cores.  The energy $E_{iz} = E_i \cos^2(\theta_i) = \hbar^2 k_{iz}^2/2m$ is the incident energy associated with motion normal to the surface for the given incident angle $\theta_i$.  In general usage $E_{iz}$ is often corrected by the Beeby correction \cite{Beeby} in which the normal energy is enhanced by the depth $D$ of the attractive adsorption well according to $E^\prime_{iz} = E_{iz} + D$.  However, for the alkali metals considered here this has a negligible effect
because the well is very shallow,
for example the Zaremba and Kohn physisorption theory \cite{Zaremba} gives for the
$^4He$-Cs surface potential $D=0.36$ meV, resulting from the sum of the lowest bound state 
energy $|E_0| = 0.13$ meV and the zero-point energy of $2.3$ meV.

Bulk electronic properties of alkali metals can often be approximated as those of a free electron gas, and He atom scattering gives evidence that near the surface the electrons may also be approximated by a free electron gas.  The evidence is that He atom scattering from all alkali surfaces thus studied exhibits only a specular diffraction peak and that all other diffraction peaks have negligible intensity.  This implies that the electronic density  near the surface is very smooth, and justifies the use here of a free electron gas DOS for the region near the surface of alkali metals.
Alkali metal crystals have body centered cubic (BCC) structure, and this holds true in the layer-by-layer growth.
For all of the alkali metals studied here the layer-by-layer growth builds up also with a BCC structure regardless of the crystal structure or crystalline face orientation of the growth substrate.

\begin{figure}
\begin{center}
\includegraphics[width=5.0in]{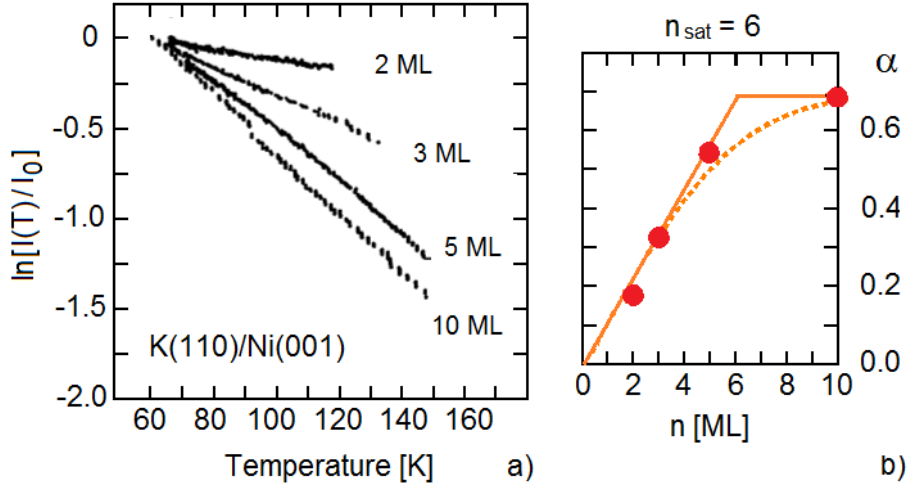}
\end{center}
\caption{a) The HAS specular reflectivity as a function of the surface temperature, normalized to the extrapolated $T = 0$ value, for 2, 3, 5 and 10 ML films of K(110) on Ni(111) (adapted from ref. \cite{Layer9}). This is plotted on a logarithmic scale in order to show the approximate linearity 
 of the Debye-Waller exponent with temperature.  Data from ref.~\cite{Layer9}  b) The average logarithmic slope of $I(T)/I_0$, expressed by the dimensionless constant  $\alpha$ of Eq. (\ref{alpha}),  increases linearly (full line) with the film thickness up to saturation at 
$n=n_{sat} = 6$. The broken line is an analytical fit (see text).}
\label{FigK-Ni}
\end{figure}

An example of the alkali atom experiments is exhibited in Fig.~\ref{FigK-Ni} a) which shows, for potassium overlayers on Ni(001), the specular He atom scattering Debye-Waller exponent  $\ln[I(T)/I_0]$ is a nearly linear function of $T$ at small monolayer (ML) numbers \cite{Layer9}.  Interestingly, the slope increases linearly with ML number $n$ for small ML numbers as shown in Fig.~\ref{FigK-Ni} b) in terms of the dimensionless quantity $\alpha$ defined below in Eq.~(\ref{alpha}), but eventually the slope saturates for ML numbers greater than $n_{sat}$.  In this case it is for ML number $ n_{sat} = 5$.  For thicker films, which for K(110)/Ni(111) were measured  up to $n>10$, the slope remains the same and may be regarded as that of semi-infinite K(110).

The density of states for a two-dimensional nearly free electron gas is given by $m_e^*/\pi \hbar^2$.
The experimentally observed linear behavior exhibited by
successive layers of
K in Fig.~\ref{FigK-Ni}b) at small ML numbers strongly suggests that in this region each monolayer contributes independently to the density of states at the Fermi surface.
Theoretical support of this supposition comes from recent  calculations of the band structure of free-standing thin films of alkali metals. \cite{Bernasconi}
The DOS can be  calculated readily from the band structure.  As an example, for a free standing film of up to 11 layers of Cs the band structure shows one parabolic quasi-free electron band per layer, each one contributing the same Fermi level DOS per unit surface area equal to  $m_e^*/\pi \hbar^2$. \cite{Bernasconi}

This linear increase of the DW slopes shown in Fig.~\ref{FigK-Ni}b) implies
that in Eq.~(\ref{Eq1})
one should use $\mathcal{N}(E_F) = \mbox{min}(n,n_{sat}) m_e^* a_{c}/\pi \hbar^2$
to indicate that in a 2D electron gas representation the layers deeper than $n=n_{sat}$ do not contribute to the electron-phonon interaction as probed by He atoms at the surface.
Then $\lambda$ derived from Eq.~(\ref{Eq1}) for any thickness $n$ becomes
\begin{eqnarray} \label{alpha}
\lambda_{HAS} ~=~ \frac{\pi}{2 \overline{n}} \alpha  ~~~; ~~~\alpha ~\equiv~ \frac{\phi~ \ln[I(T_1)/I(T_2)]}{a_{c} k^2_{iz} k_B (T_2-T_1)}~ ,   ~~~
\overline{n} \equiv \mbox{min}(n, n_{sat})~,
\end {eqnarray}
where the index $\overline{n}$ is used to indicate that the measured specular intensities refer to a film of arbitrary thickness $n$.
Here $a_{c}$ is the area of the surface unit cell and
$T_1$ and $T_2$ are any two temperatures in the region where $2W({\bf k}_f, {\bf k}_i;T)$ is linear in $T$. Because the dimensionless function $\alpha$, derived from the HAS data and plotted in Fig.~\ref{FigK-Ni}b), initially grows linearly with n, $\lambda$ consequently has about the same value for $n \le  n_{sat}=6$.  The average over the three ($n=2,3,5)$ films produces the value
$\lambda_{HAS} = 0.16 \pm 0.02$. The actual dependence of $\alpha$ and $n$ can be estimated
by considering that the thermal vibrations of deep layers, inducing the surface charge density
fluctuations measured by HAS, rapidly decay for increasing $n$. A Gaussian approach of $\alpha$
to saturation like $\alpha = \alpha_{sat} (1 - \exp{(- n^2 / n^2_{sat}}))$ with 
$\alpha_{sat} = 2 \lambda_{HAS} n_{sat} / \pi$ is suggested (broken line in Fig. 1 (b)); at 
$n=n_{sat}$, this yields $\lambda_{HAS} = 0.146$, consistent with the above average value.   

After saturation, i.e., for $n>6$ for the present case of potassium, the slope of the Debye-Waller plot no longer depends on the number of monolayers so it is appropriate to use in   Eq.~(\ref{Eq1}) the DOS for a free electron gas,  $\mathcal{N}(E_F) = 3 Z m^*_e/ \hbar^2 k_F^2 $, where $Z$ is the number of  electrons provided by each substrate metal atom and $k_F$ is the Fermi wave vector.  Using $Z=1$ and  0.27~\AA$^{-1}$ for $k_F$ \cite{kF-layer} one obtains the same value as found above for $n \le n_{sat}$, i.e.,   $\lambda_{HAS}=0.16$, which compares favorably with the known tabulated bulk values which range from 0.11 \cite{AllenT} to 0.13$\pm$0.04 \cite{Grimvall}.

\begin{figure}
\begin{center}
\includegraphics[width=5.0in]{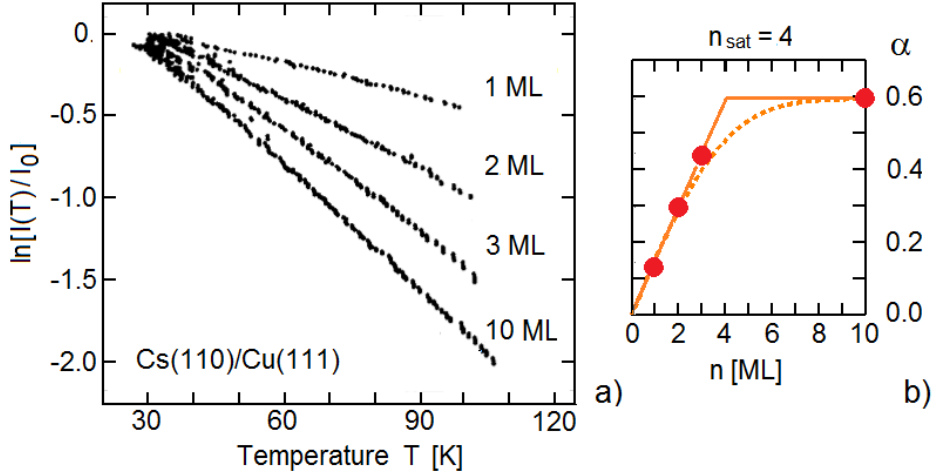}
\end{center}
\caption{Similar to Fig.~\ref{FigK-Ni}, but for Cs(110) films on Cu(111). Data from ref.~\cite{Layer9}.}
\label{FigCsCu}
\end{figure}

Similar measurements have been made for Cs(110)/Cu(111) as shown in Fig.~\ref{FigCsCu}.  It is seen that for this system saturation also begins after $n_{sat} = 4$.  In this case the value of $\lambda_{HAS}$ produced by the first three layers turns out to be 0.22, if $\phi = 2.14$ eV is used
from Michaelson´s compilation  \cite{k} or 0.18 if we take $\phi = 1.81$ eV  as found in the 
more recent compilation by Kiejna and Wojciechowski. \cite{81Kie}
It should be noted that the work function is known to vary in thin films as a function of thickness \cite{Qi}. Its oscillations are however small as compared to the present level of accuracy, and will be considered only in presence of important quantum size effects, as discussed below for Pb
films, while for alkali films the bulk values of $\phi$ are considered a sufficient approximation.  
At saturation for $n>4$, again using the three-dimensional free electron DOS in Eq.~(\ref{Eq1}), the value produced is also 0.18 (0.15) which compares favorably with the tabulated bulk values of 0.15 \cite{Grimvall} and 0.16. \cite{AllenT}
The values of $\lambda_{HAS}$ for $n \le n_{sat} $ for these and  all the other alkali metal systems studied here, together with more details, are contained in Table~\ref{layertable}.
The ability to study Cs with He atom scattering is interesting as it is one of the few elements that is difficult to study in the bulk with neutron scattering because of the very high neutron capture cross section of the Cs nucleus.  He atom scattering at a Cs surface is not hampered by such problems.

\begin{figure}
\begin{center}
\includegraphics[width=5.0in]{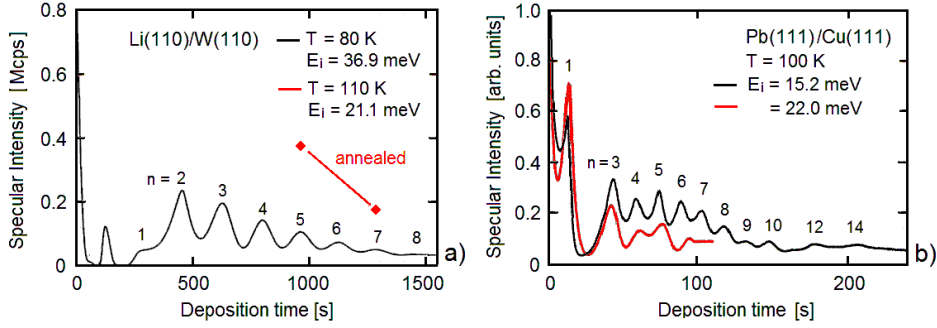}
\end{center}
\caption{a) Specular He atom scattering intensity as a function of deposition time (proportional to coverage) for Li(110) deposited on a W(110) substrate (from Ref.~\cite{Layer2}).  The ML numbers are indicated.  The specular intensity saturates above $n=8$.
In this system there is a small maximum at 0.5 coverage, while the monolayer at $n = 1$ is rather disordered because of the conflicting periodicities of W(110) and the Li (110) layer. 
Measurements on annealed samples (red lozenges) give  substantially larger specular intensities due to defect bleaching, yielding however about the same $\lambda$.  
b) The growth of Pb films on Cu(111), here shown for comparison and observed at two different He incident energies, shows a similar decay of the specular intensity for increasing layer number (though with oscillations due to the quantum size effect \cite{l}), with the formation of a wetting layer at $n = 1$, a disordered second layer, followed by a regular growth for $n \geq 3$ and saturation at $n = 9$ (from ref. ~\cite{Braun}). }
\label{FigLi-W}
\end{figure}

In addition to Debye-Waller plots such as Figs.~\ref{FigK-Ni}a) and~\ref{FigCsCu}a) another important type of measurement which is often carried out is the monitoring of layer-by-layer growth from oscillations in the specular intensity as a function of deposition time, taken at a fixed temperature.  An example of Li(110) monolayers deposited on a W(110) substrate at $T=80$ K and $E_i = 37$ meV is shown in Fig.~\ref{FigLi-W}.  Such measurements provide another method for obtaining the $\lambda$ value for $n \leq n_{sat}$ by comparing intensities of the oscillation maxima (which occur at full ML coverage) for different ML numbers.  Again, this possibility depends on the known linear  behavior in temperature of the logarithmic Debye-Waller plots.
To show this consider comparing the intensity maxima for the $n$ and $n+\ell$ monolayers (with both $n$ and $n+\ell<n_{sat}$) denoted by $I_{n}(T)$ and $I_{n+\ell}(T)$, respectively.
In general, if layer-by-layer growth plots such as Fig.~\ref{FigLi-W} are carried out at two different temperatures $T_1$ and $T_2$, with both of these temperatures within the range  over which the DW plots are linear in $T$, Eq.~(\ref{alpha}) can be expanded to produce a value of $\lambda$ from the following combination
\begin{eqnarray} \label{lambda2}
\lambda ~=~\frac{(n+\ell) \lambda -n \lambda}{\ell}   ~=~ \frac{\pi \phi}{2 \ell a_{c} k_{iz}^2 k_B (T_2-T_1)}
\left[ \ln\left( \frac{I_{n+\ell}(T_1)}{I_{n+\ell}(T_2)} \right) ~-~ \ln\left( \frac{I_{n}(T_1)}{I_{n}(T_2)} \right)  \right]
\nonumber \\
 ~=~ \frac{\pi \phi}{2 \ell a_{c} k_{iz}^2 k_B (T_2-T_1)}
\left[ \ln\left( \frac{I_{n+\ell}(T_1)}{I_{n}(T_1)} \right) ~-~ \ln\left( \frac{I_{n+\ell}(T_2)}{I_{n}(T_2)} \right)  \right]
~,
\end {eqnarray}
provided that $n+1 \leq n_{sat} ~\mbox{and}~ \ell +n \leq n_{sat}$.

\begin{figure}
\begin{center}
\includegraphics[width=4.0in]{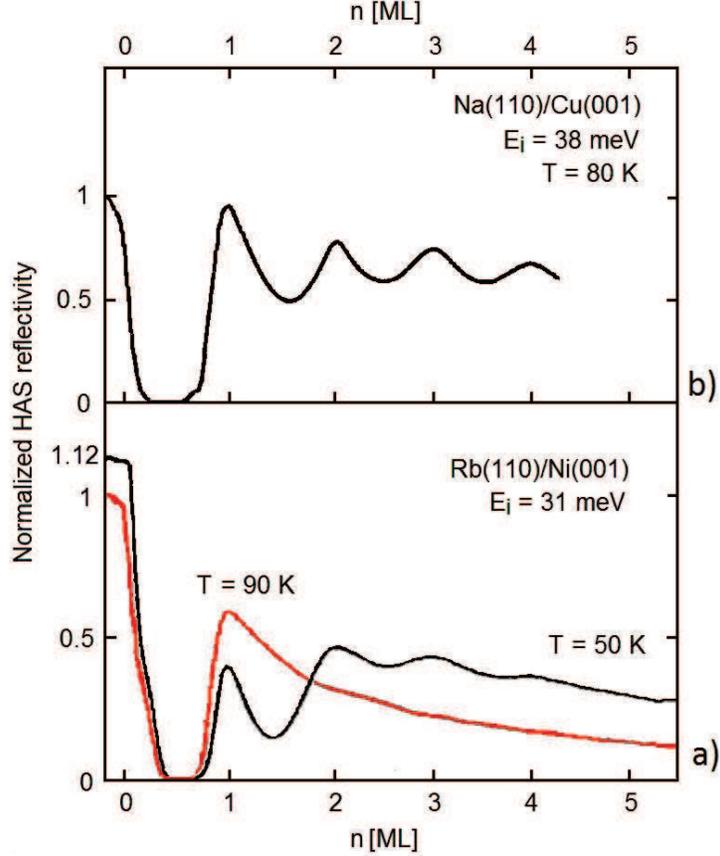}
\end{center}
\caption{
a)~Similar to Fig.~\ref{FigLi-W} but for Rb(110) monolayers on a Ni(001) substrate, showing measurements at two different temperatures of 50 and 90 K as marked. From ref.~\cite{Layer5} b)  For Na(110) monolayers on a Cu(001) substrate at $T=80$ K, showing clear saturation behavior.   From ref.~\cite{Layer8}.}
\label{FigNaRb}
\end{figure}

An even simpler expression can be obtained upon recognizing that the extrapolation of the specular intensity to $T = 0$ becomes independent of layer number, or can be normalized to the same value as shown in Figs.~\ref{FigK-Ni} and~\ref{FigCsCu}.  Thus, if the intensity at $T_1$ is taken to be that obtained by extrapolation to $T \longrightarrow 0$, Eq.~(\ref{lambda2}) simplifies to
\begin{eqnarray} \label{lambda1}
\lambda ~=~
 \frac{\pi \phi}{2 \ell a_{c} k_{iz}^2 k_B T}
\ln\left( \frac{I_{n}(T)}{I_{n+\ell}(T)} \right)
~,
\end {eqnarray}
as long as the temperature $T$ is within the region for which $2W$ is linear in $T$.
A distinct advantage of using the layer-by-layer growth methods of either Eq.~(\ref{lambda2}) or~(\ref{lambda1}) is that  even if the growth curve is measured at only a single temperature, such as  the case for Li(110) in Fig.~~\ref{FigLi-W},
all combinations of two different peaks provide distinct values of $\lambda$, and this gives sufficient numbers of values for which a standard deviation of error can be evaluated.  If the growth plots are carried out at two or more different temperatures, such as the case for Rb(110)/Ni(001) in Fig.~\ref{FigNaRb}a),  Eq.~(\ref{lambda2}) can be used and this provides even greater numbers of distinct values for $\lambda$.

The HAS specular intensity patterns for Li(110) on a W(110) substrate of Fig.~\ref{FigLi-W}a) and Na(110)/Cu(001) in Fig.~\ref{FigNaRb} b), both deposited at a temperature of 80 K, show saturation behavior.  For Li(110)/W(110)  the saturation occurs after $n=8$ and using Eq.~(\ref{lambda1}) the average over $n=2 \longrightarrow 8$ gives $\lambda_{HAS} =0.54 \pm 0.08$, while the more restricted average over $n=2 \longrightarrow 5$ gives $\lambda_{HAS} =0.47 \pm 0.11$.  By comparison the bulk value ranges from $\lambda=0.35$\cite{AllenT} to 0.40~\cite{Grimvall}.  
The derivation of $\lambda$ from the growth slope could be misleading because the contribution 
of the 
e-ph interaction could be obliterated by that because of the accumulation of defects. This is not so however. The available data on the specular intensities of annealed 5 and 7 ML Li(110) films on W(110) (red lozenges in Fig. 3 a)) \cite{ Layer2} , where defects are nominally absent, yield 
$\lambda_{HAS} = 0.46 \pm 0.05$, in close agreement with the values obtained from the growth slopes. This suggests that the contributions of defects and e-ph interaction to the reduction of the specular intensity with increasing $n$ are proportional before saturation, so that the formation of defects has no effect on the intensity ratios of Eqs. (3) and (4).  
The pattern of Na(110) growth in Fig.~\ref{FigNaRb} suggests a very regular growth, and the average over $n=2 \longrightarrow 4$ gives  $\lambda_{HAS} =0.17 \pm 0.03$ as compared to  $0.16$ for bulk Na~\cite{Grimvall} and 0.24 for the surface of a Na quantum well.~\cite{Layer4}
Such a regularity is apparently a property of simple metals. For comparison the layer-by-layer growth of Pb films on Cu(111) (Fig.~\ref{FigLi-W} b)) shows an oscillating decay of the He specular scattering which has been ascribed to the quantum size effect (QSE) \cite{l}.
Nevertheless there is a clear saturation at $n = 9$, which allows  to derive from the interval
$n = 3 \rightarrow 9$ 
and  for $E_i = 15.2$ meV (Fig. 3 b)) 
a mass-enhancement factor $\lambda_{HAS} = 1.18$. There are however some oscillations of 
$\lambda_{HAS}$ when referred to different intervals: e.g., with $E_i = 15.2$ meV and for  
$n = 3 \rightarrow 8$  it is found $\lambda_{HAS} = 0.93$ while for  $n = 3 \rightarrow 7$ 
 it is $\lambda_{HAS} = 1.10$; on the other hand for $E_i = 22.0$ meV and $n = 3 \rightarrow 7$  one finds $\lambda_{HAS} = 0.90$ (see Table~\ref{layertable}). As shown below the QSE 
causes oscillations of  $\lambda_{HAS}$  with the film thickness.

With regards to Rb(110)/Ni(001) in Fig.~\ref{FigNaRb}a) $\lambda_{HAS}$ can be derived from the DW temperature dependence over the range of $n = 2-5$ and for the two different temperatures of 50 and 90 K using both Eqs.~(\ref{lambda2}) and~(\ref{lambda1}).
The average is $\lambda_{HAS} = 0.19 \pm 0.06$ as compared with the standard bulk value of 0.15.~\cite{AllenT} This bulk value is smaller, and also smaller than the values extracted via Eq.~(\ref{lambda2}) from the DW plot slopes~\cite{Layer8} for $n > 2$ which are 0.19 at 50 K and 0.22 at 90 K.  This suggests that a non-negligible contribution to the decrease of the specular intensity comes from the increase of defects with temperature, as indicated also by the small layer-by-layer oscillations. On the other hand the lack of large oscillations, ostensibly due to defects, does not seem to affect the value of $\lambda_{HAS}$ which encompasses the electron-phonon interaction of the whole film as probed by HAS.

It is important to point out another advantage of obtaining $\lambda$ from the layer-by-layer growth plots, which is that effects due to disorder and defects tend to cancel.
This can be seen from the first line of Eq.~(\ref{lambda2}); if at $T_1$ and $T_2$ the additional attenuation of the $n$-th and $n + \ell$-th peaks due to static disorder is identical, then it cancels.
This is evident because the attenuating effect of disorder is usually expressed as a multiplicative factor (sometimes called a characteristic function \cite{Beckmann}) applied to each diffraction peak, regardless of whether the disorder is due to small displacements of the surface or due to defects, 
and in the intensity ratios appearing in Eq.~(\ref{lambda2}) or~(\ref{lambda1}), such factors cancel.  Clearly in the growth spectra such as in Figs.~\ref{FigLi-W} and~\ref{FigNaRb} both disorder and changes in the electron-phonon interaction must play a role in the fact that the layer-by-layer peaks gradually become less pronounced.  In some cases the specular intensity reaches a nonzero saturation value, such as for Li(110) in Fig.~\ref{FigLi-W} or Na(110) in Fig.~\ref{FigNaRb}b), while in other cases the specular intensity appears to continue decreasing with large coverage such as for Rb at $T=90$ K in Fig.~\ref{FigNaRb}.  For the systems which exhibit this saturation behavior with coverage it appears reasonable to assume that disorder attenuation is less important and the gradual disappearance of the layer-by-layer peaks is indicative of the influence of the electron-phonon interaction.  For the systems that do not exhibit such saturation with coverage the disorder due to defects in the growing overlayer is surely playing a much larger role, but these systems are still amenable to analysis using Eq.~(\ref{lambda2}) or~(\ref{lambda1}) because, as mentioned above, the effects of disorder tend to cancel out of the ratios of the specular intensity for the same $n$ taken at two different temperatures.

\begin{figure}
\begin{center}
\includegraphics[width=6.0in]{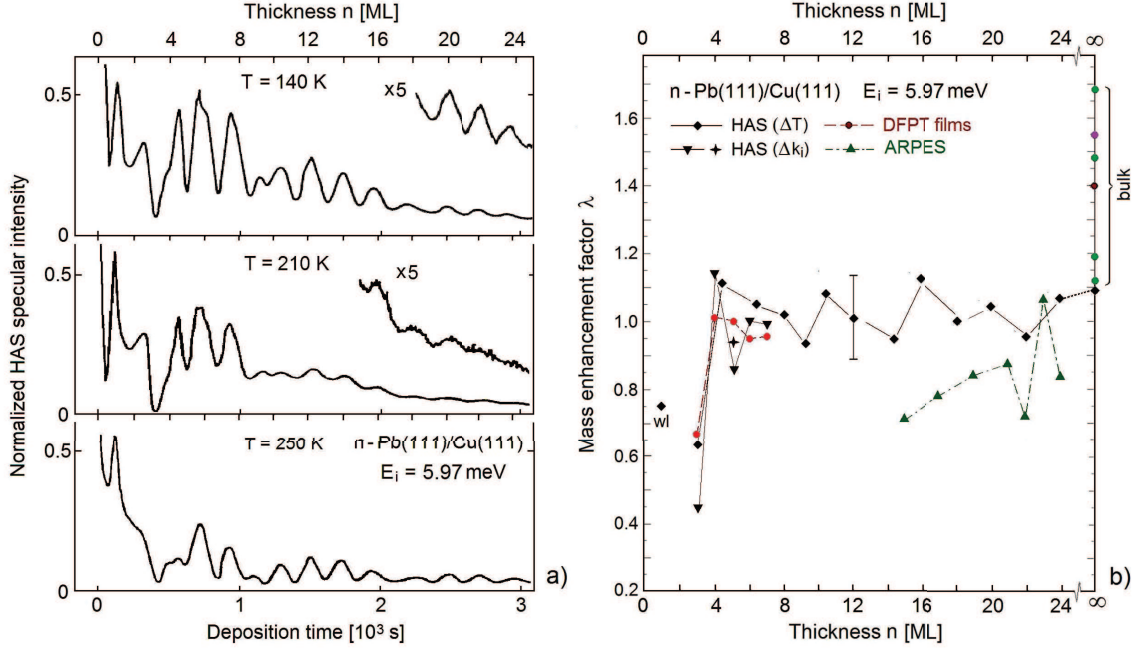}
\end{center}
\caption{a) Bilayer-by-bilayer growth curves for Pb(111) deposited on Cu(111) at temperatures of 140, 210 and 250 K as measured with HAS at the incident energy of 5.97 meV~\cite{l,GZhang}.  b) Values of $\lambda_{HAS}$ extracted from the growth curves shown in panel a) (filled diamonds connected by a solid line).  The isolated diamond at $n=1$ marked ``wl" is from the wetting layer.  The downward-pointing filled triangles at small $n$ are values of $\lambda_{HAS}$ extracted from the energy dependence of the Debye-Waller factor (Fig.~\ref{FigLi-W} b))~\cite{Braun}. The point marked by a cross is from energy dependent measurements made at $n=5$ taken in conjunction with the temperature-dependent data. \cite{GZhang}
Values of $\lambda$ calculated with density functional perturbation theory for thin films are shown as filled circles connected by a dashed line~\cite{Skl}.
The upward-pointing triangles connected by dash-dotted lines are ARPES measurements. \cite{Zhang-05}
A collection of values for bulk lead taken from the literature \cite{Grimvall,AllenT} (full circles at $n \rightarrow \infty$)  are shown for comparison.
 }
\label{FigPb}
\end{figure}

The comparatively simple physics concerning the electron-phonon interaction in alkali ultrathin films cannot be straightforwardly extended to other metals such as, for example, lead.  The evolution of the HAS specular peak intensity of Pb(111) ultrathin films has been extensively studied as a function of thickness in the layer-by-layer growth regime at different temperatures~\cite{l,GZhang,Braun}. Pb films grown on Cu(111) have in common with the alkali films discussed above the fact of being considerably softer than the substrate, which practically restricts the phonon spectrum involved in electron-phonon interaction to that of the film on a rigid substrate~\cite{Skl}. However the HAS normalized specular intensity of Pb films measured during growth as a function of thickness at three different temperatures shown in Fig.~\ref{FigPb}a) exhibits, over a range up to about 30 Pb layers, rapid oscillations indicating a bilayer-like growth.  These growth peaks are further modulated by an envelope of longer-period oscillations, having a length of about 8 layers, which have been convincingly interpreted as quantum size effects~\cite{l}.  The comparatively large interlayer distance contraction affecting the topmost bilayer~\cite{Chulkov,Yndurain} causes the split-off of a surface phonon branch above the bulk phonon spectrum.  This explains the bilayer growth mode and suggests that the oscillations of the HAS DW exponent essentially depend on those of the top bilayer 2D electron gas as induced by phonons of the entire film.  In this case, an examination of the logarithmic DW plots, similar to Figs.~\ref{FigK-Ni} and~\ref{FigCsCu} above, using the three available temperatures of Fig.~\ref{FigPb}a) indicates that
$n_{sat} = 9$ is  a reasonable approximation.

The values of $\lambda_{HAS}$ derived from Eq.~(\ref{lambda2}) for the three temperatures of Fig.~\ref{FigPb}a) are plotted, as a function of the nominal thickness layer number $n$, as filled diamond points connected by solid lines in Fig.~\ref{FigPb}b).
Values of the work function are taken from measurements of $\phi$ as a function of Pb layer thickness $n$ on a Si(111) substrate~\cite{Qi}.
Meaningful values start from $n = 3$, because the first monolayer of Pb on Cu(111) actually plays the role of a wetting layer, and the second ML film is disordered~\cite{l,GZhang}.  However, the value of $\lambda_{HAS}$ extracted from the $n=1$ wetting layer is shown as an isolated diamond marked as ``wl".
The vertical bar represents the standard deviation uncertainty. 
No Beeby correction for the well depth was made in the calculations, but energy dependent measurements of the specular intensity indicate that this is a reasonable approximation as discussed below.

The form of Eq.~(\ref{Eq1}) above makes it clear that $\lambda_{HAS}$ can also be obtained from measurements of the DW exponent $2W(E_{iz}, T)$ at fixed temperature, but for two different incident energies.  However, such measurements require a correction for the change in intensity of the incident beam as a function of energy.  In a series of independent measurements this problem was addressed by changing the source temperature at constant pressure, in which case the supersonic beam flux to a good approximation varies inversely in proportion to $k_i$.~\cite{Braun}
For such a measurement at two different energies the mass enhancement factor $\lambda_{HAS}$ is again given by Eq.~(\ref{alpha}) with the modification that the intensity must be corrected by a factor of $k_i$ and $\alpha$ is replaced by
\begin{eqnarray}  \label{alpha2}
\alpha ~=~ \frac{\phi}{a_c k_B T} ~ \frac{\Delta \ln[k_i I(E_{iz}, T))]}{\Delta k_{iz}^2}
~,
\end {eqnarray}
where $\Delta \ln[k_i I(E_{iz}, T))]$  and  $\Delta k_{iz}^2$ are the differences between the respecitve quantities at the two  incident energies.
Measurements of growth curves of Pb on Cu(111) ) made at $T  = 100$ K and two incident energies with wavevectors of 5.4 and 6.5 \AA$^{-1}$ (Fig.~\ref{FigLi-W} b)) show bilayer growth of up to $n > 10$ for the lower energy and $n = 7$ for the larger energy~\cite{Braun}.
Plotted in Fig.~\ref{FigPb}b) as downward-pointing triangles are the values of $\lambda_{HAS}$ extracted from this data, and the values are found to be in excellent agreement with those obtained from the temperature dependence at fixed incident energy.  It is important to note that the possible dependence on the physisorption well depth $D$ through a Beeby correction cancels out in the energy difference $\Delta k_{iz}^2$ of Eq.~(\ref{alpha2}).  Thus the good agreement between the energy-dependent and temperature-dependent results shown in Fig.~\ref{FigPb}b) indicates that the Beeby correction for the well depth is unimportant in these Pb layer systems, and also indicates that the incident beam intensity correction for the energy is a reasonable approximation.

In Fig.~\ref{FigPb}b) the isolated point at $n=5$ marked with a cross is from the single measurement of the incident energy dependence of the DW exponent taken during the same experiment in which the temperature-dependent data were obtained.~\cite{GZhang}  The agreement with the temperature-dependent measurements is quite good.

These extracted $\lambda_{HAS}$ values are  compared in Fig.~\ref{FigPb}b) with previous measured values for thick films of 15 to 24 layers using angle-resolved photoemission spectroscopy (ARPES)  which lie in the range 0.7 - 1.05~\cite{Zhang-05} as shown as upward-pointing triangles connected by dash-dotted lines.  Also shown as filled circle points are values for thin films calculated using density functional perturbation theory \cite{Skl}, and
older bulk values in the literature which range from 1.12 to 1.68.~\cite{Grimvall,AllenT} It is 
interesting to note that density functional theory  calculations for free-standing films (4-10 ML) by Sklyadneva et al. 
\cite{13Skl} yield values of $\lambda$ oscillating between 1.1 and 1.3 without spin-orbit interaction,
while its inclusion raises $\lambda$ to within 1.5 and 2.0, with more prominent quantum-size
oscillations. Although the latter results do not agree so well with the present values of 
$\lambda_{HAS}$, they reveal the relevant contribution to electron-phonon interaction of 
spin-orbit interaction. 

\vspace{1cm}
\begin{table}
\caption{
The mass enhancement factor $\lambda_{HAS}$ derived from the specular intensities of HAS experiments on alkali metal  and Pb layers deposited on different close packed metal substrates.  Also shown are the relevant experimental parameters as well as values of $\lambda$ from other sources as cited.  The brackets $\langle \rangle$ signify an average taken over all $n$ specular maxima in the layer-by-layer growth spectra subject to $n \le n_{sat}$.}
\vspace{1cm}
\centering
\begin{tabular}{|c||c|c|c|c|c|c|c|}
  \hline      Growing & $ \left\langle \ln \left[I_{n} / I_{n+1}  \right] \right\rangle $  & $T$
   & $ k_{iz}^2  $ & $ \phi $  & $n$ & $ \lambda_{HAS}$ & $\lambda$~(other     \\
   layers  &     & $[\mbox{K}]$& $[\mbox{\AA}^{-2}] $ & $\left[ \mbox{eV} \right]$ & $n \leq n_{sat}$ & & sources)  \\
  \hline
  \hline  Li(110)/     &  0.261   &80 &  34.9  &
    2.32 \cite{Layer1} &2-5 & 0.47 $\pm$ 0.11  & 0.40  \cite{Grimvall} \\
   W(110) \cite{Layer2} & &110 & 22.3 & &5-7 &0.46 $\pm$ 0.05 & 0.35 \cite{AllenT}\\
  \hline  Na(110)/     &  0.119   &80 &  35.9  &
    2.75 \cite{k} &2-4 & 0.17 $\pm$ 0.03  & 0.16  \cite{Grimvall} \\
    Cu(001) \cite{Layer5}
    & & & & & & & 0.24 \cite{Layer4} \\
  \hline  K(110)/     & 0.206   &100 &  27.1  &
    2.30 \cite{k} &2-6 & 0.16 $\pm$ 0.02  & 0.13  \cite{Grimvall} \\
       Ni(001) \cite{Layer9} & & & && & & 0.11 \cite{AllenT} \\
  \hline  Rb(110)/     &  0.296   &90 &  29.7  &
    2.16 \cite{k} &2-5 & 0.19 $\pm$ 0.06   & 0.16  \cite{Grimvall}\\
      Ni(001) \cite{Layer9} &    &  &  & &  &  &  0.15 \cite{AllenT}  \\
  \hline  Cs(110)/     &  0.346   &75 &  26.9  &
    2.14 \cite{k} & 1-4 &  0.22 $\pm$ 0.02 & 0.15  \cite{Grimvall} \\
     Cu(111) \cite{Layer9}   & & & & 1.81 \cite{81Kie} & & 0.18 $\pm$ 0.02 & 0.16 \cite{AllenT} \\
  \hline  Pb(111)/     &  1.31   &100 &  29.2  &
    4.1 \cite{Qi} & 3-9 & 1.18  & 1.12-1.68 \cite{Grimvall,AllenT} \\
     Cu(111) \cite{l}   &0.86 & & 29.2&   & 3-8 &0.93 & 0.95 \cite{Benedek-14} \\
             &0.97 & & 42.3&   & 3-7 & 0.90&  0.7-1.05 \cite{Zhang-05}  \\
\hline
\end{tabular}
\vspace{1cm}
\label{layertable}
\end{table}

It is important to note that in Table I the values of $\lambda_{HAS}$, directly derived from 
experiment with no specification besides the layer number, the film work function, the surface
atom density and the He atom incident momentum, are compared to available {\it bulk} values of 
$\lambda$, either obtained from other experimental methods or theoretically. For alkali films
(except Na) $\lambda_{HAS}$ turns out to be systematically larger than the bulk $\lambda$
by more than 20 $\%$. There are different reasons for that worth being elucidated with new HAS
experiments and  a theoretical refinement. The first possible reason is the intrinsic enhancement
of e-ph interaction expected at metals. \cite{plummer} In supported films $\lambda$ may also
differ from the surface value for the corresponding semi-infinite crystal due to different electronic
structure and lattice dynamics of the substrate. In comparing the measured $\lambda_{HAS}$ with 
bulk values it is implicitily assumed that the substrate is inert, in the sense that its vibrations are
sufficiently decoupled from those of the film so as to be unable to modulate the film surface 
charge density probed by HAS, and its Fermi level electronic states do not contribute appreciably 
to the surface charge density. This assumption is likely to work well for alkali films, whereas
for a few MLs of Pb films on Cu recent embedded atom calculations including the substrate 
dynamics \cite{rusina} reveal changes with respect to rigid-substrate calculations, especially 
for the polarization of the interface phonon which plays a relevant role in quasi-2D superconductivity
of Pb ultrathin films \cite{Skl}; the fact that in the case of Pb films on Cu the substrate $\lambda$
is much smaller than that of the film may compensate the surface enhancement effect.

Finally, the values of $\lambda_{HAS}$ derived from the reflectivity of growing films, 
Eq.~(\ref{lambda1}),
may be affected by a positive systematic error due to the increasing number of steps during growth, 
which yields a faster decrease of reflectivity. However, the reflectivity maxima for increasing 
$n$ measured on annealed samples (an example is shown in Fig. 3a)) turn out to be much larger 
than  but proportional to those for a continuous growth with no annealing, so that no effect is expected
on $\lambda_{HAS}$ from Eq.~(\ref{lambda1}). In conlcusion, the observed excess of 
$\lambda_{HAS}$ in
alkali films with respect to bulk $\lambda$ may be considered as a genuine surface enhancement,
which may however be corrected by substrate effects  as is likely to occur in Pb films on Cu.

In this Letter we have shown that through the use of He atom scattering to monitor the temperature dependence of layer-by-layer growth, the electron-phonon coupling constant $\lambda$ can be obtained for systems of metal atomic layers grown on different close packed metal substrates.  The method is based on earlier work  which showed that the Debye-Waller exponent is directly proportional to $\lambda$ for simple metal crystal surfaces, as exhibited in Eq.~(\ref{Eq1}).  For the case of the multiple atomic monolayers examined here, namely layers of alkali metals and Pb, it is shown that for small monolayer numbers ($n \le n_{sat}$) the three-dimensional density of electronic states at the Fermi surface which appears in Eq.~(\ref{Eq1}) must be replaced by the two-dimensional density of states contributed by all of the $n \le n_{sat}$ layers.  For monolayer thicknesses with $n>n_{sat}$ the 3D DOS is appropriate.  The use of the 2D density of states is supported, at least for alkali metals, by recent calculations of the band structure for free-standing thin films of Cs, where it is found that each successive layer independently contributes a quasi-free parabolic band, implying additivity of the DOS at the Fermi level \cite{Bernasconi}.  This property is empirically confirmed in the data which shows that for both alkali metal and Pb overlayers with $n \le n_{sat}$ the slope of the logarithmic DW versus $T$ plots is linearly dependent on the ML number $n$.  The values of $\lambda$ obtained from the temperature dependence of the DW factors, for both small ML values and for films sufficiently thick to be considered as semi-infinite, compare favorably with known measured bulk values or with previous calculated values.

The fact that the DW exponent is linear in temperature, as shown in Eq.~(\ref{Eq1}), and also linear in ML number for $n \le n_{sat}$ provides a second way of extracting values of $\lambda$.  This is through comparing ratios of intensities of a given  diffraction peak (usually the specular), measured at full monolayer coverage, but with two different ML numbers and at two different temperatures.  Compared to the method of extracting $\lambda$ from the slopes of the temperature-dependent Debye-Waller plots, this alternative method has the advantage that attenuation of intensity due to static defects tends to cancel out.  It also provides a large number of different ML number combinations, each of which produces a value of $\lambda$, which gives sufficient numbers for providing statistical standard deviations.  For the Pb layer-by-layer growth as well as for the alkali metals the alternative method produced values of $\lambda$ that were in reasonable agreement with the values obtained from the logarithmic Debye-Waller plots, and which compared favorably  with the known bulk values.

The reasonable agreement between the surface $\lambda$ values obtained in this work compared to the known bulk values indicates that He atom scattering can measure electron-phonon interaction properties of surfaces, not only for the simple metals for which Eq.~(\ref{Eq1}) was originally derived, but also for more complicated systems such as the metallic layer-by-layer growth systems examined here.  In particular, the electron-phonon coupling strength can be measured as a function of the thickness of the film.

~\\
Acknowledgment: We would  like to thank Profs. M. Bernasconi, E. V. Chulkov and P. M.
Echenique (DIPC) for helpful discussions.
This work is partially supported by a   grant with Ref.
FIS2014-52172-C2-1-P from the Ministerio
de Economia y Competitividad (Spain).






\newpage

\begin{thebibliography}{199}


\bibitem{Reviews2} Goodman, F. O.; Wachman, H. Y. {\em Dynamics of Gas-Surface Scattering}; Academic Press: New York; 1976.

\bibitem{Reviews} Hulpke (Ed.), E. {\em Helium Atom Scattering from Surfaces}; Springer Series in Surface Sciences {\bf 27}; Springer Press: Heidelberg; 1992.

\bibitem{Farias} Far\'ias, D.; Rieder, K.-H. Atomic Beam Diffraction from Solid Surfaces. 
{\em Rep. Prog. Phys.} {\bf 1998}, 61, 1575-1664.

\bibitem{Disorder}  Poelsema, B.;  Comsa, G. {\em Scattering of Thermal Energy Atoms from Disordered Surfaces}; Springer Tracts in Modern Physics {\bf 115}; Springer Press: Berlin; 1989.

\bibitem{Doak}  Brusdeylins, G.; Doak, R. B.; Toennies, J. P. Observation of Surface Phonons in Inelastic-Scattering of He Atoms from LiF(001) Crystal-Surfaces. 
{\em Phys. Rev. Lett.} {\bf 1980}, 44, 1417-1420.


\bibitem{Bertino}  Bertino, M.; Ellis, J.; Hofmann, F.; Toennies, J. P.; Manson, J. R. High-Resolution Helium Scattering Studies of Inelastic Interference Structures of the Frustrated Translational Mode of CO on Cu(001). {\em Phys. Rev. Lett.} {\bf 1994}, 73, 605-608.

\bibitem{CrossSections} Poelsema, B.; Comsa, G. Scattering of Thermal He from Disordered Surfaces.{\em  Faraday Discuss. Chem. Soc.} {\bf 1985}, 80, 247-256.

\bibitem{Comsa}  Kunkel, R.; Poelsema, B.; Verheij, L. K.; Comsa, G. Reentrant Layer-by-Layer Growth During Molecular-Beam Epitaxy of Metal-on-Metal Substrates. 
{\em Phys. Rev. Lett.} {\bf 1990}, 66, 733-736.

\bibitem{Skl} Sklyadneva, I. Yu.; Benedek, G.; Chulkov, E. V.; Echenique, P. M.; Heid, R.; 
Bohnen, K.-P.; Toennies, J. P. Mode-Selected Electron-Phonon Coupling in Superconducting Pb Nanofilms Determined from He Atom Scattering. {\em Phys. Rev. Lett.} {\bf 2011}, 107, 095502.


\bibitem{13Tam} Tamt\"ogl, A.; Kraus, P.; Mayrhofer-Reinhartshuber, M.;  Campi, D.; Bernasconi, 
M.; Benedek, G.; Ernst, W. E.  Surface and Subsurface Phonons of Bi(111) Measured with Helium Atom Scattering. {\em  Phys. Rev. B} {\bf 2013}, 87, 035410.


\bibitem{13Tame} Tamt\"ogl, A.; Kraus, P.; Mayrhofer-Reinhartshuber, M.;  Campi, D.; Bernasconi, 
M.; Benedek, G.; Ernst, W. E.  Surface and Subsurface Phonons of Bi(111) Measured with Helium Atom Scattering. {\em  Phys. Rev. B} {\bf 2013}, 87, 159906 (E).


\bibitem{13Kra} Kraus, P.; Tamt\"ogl, A.; Mayrhofer-Reinhartshuber, M.; Benedek, G.;  Ernst, W.
E. Resonance-Enhanced Inelastic He-Atom Scattering from Subsurface Optical Phonons of Bi(111). {\em Phys. Rev. B} {\bf 2013}, 87, 245433.




\bibitem{Benedek-14}  Benedek, G.; Bernasconi, M.; Bohnen, K.-P.; Campi, D.; Chulkov, E. V.; 
Echenique, P. M.; Heid, R.; Sklyadneva, I. Yu.; Toennies, J. P. Unveiling Mode-Selected Electron-Phonon Interactions in Metal Films by Helium Atom Scattering. {\em Phys. Chem. Chem. Phys.} 
{\bf 2014},16, 7159-7172.


\bibitem{Manson-JPCL-16}  Manson, J. R.; Benedek, G.; Miret-Art\'es, S. Electron-Phonon 
Coupling Strength at Metal Surfaces Directly Determined from the Helium Atom Scattering 
Debye-Waller Factor. {\em J. Phys. Chem. Lett.} {\bf 2016}, 7, 1016-1021.


\bibitem{17Tam} Tamt\"ogl, A.; Kraus, P.; Avidor, N.; Bremholm, M.; Hedegaard, E. M. J.; 
 Iversen, B. B.;  Bianchi, M.;  Hofmann, P.; Ellis, J.; Allison, W.; Benedek, G.; Ernst, W.E.
Electron-Phonon Coupling and Surface Debye Temperature of Bi2Te3(111) from Helium Atom Scattering. {\em Phys. Rev. B} {\bf 2017}, 95, 195401.


\bibitem{Manson-SurfSciRep}  Manson, J. R.; Benedek, G.;  Miret-Art\'es, S. Atom Scattering as a Probe of the Surface Electron-Phonon Interaction. {\em Surface Science Reports}, unpublished.

\bibitem{Beeby} Beeby, J. L. Scattering of Helium Atoms from Surfaces. {\em J. Physics C} 
{\bf 1971}, 4, L359-L362.

\bibitem{Zaremba} Zaremba, E.; Kohn, W. Theory of Helium Adsorption on Simple and Noble-Metal Surfaces. {\em Phys. Rev. B} {\bf 1977}, 15, 1769-1781.


\bibitem{Layer9} Hulpke, E.; Lower, J.; Reichmuth, A. Strain and Confined Resonances in Ultrathin Alkali-Metal Films. {\em  Phys. Rev. B} {\bf 1996}, 53, 13901-13908.

\bibitem{Bernasconi} Campi, D.; Bernasconi, M.; Benedek, G.; Graham, A. P.;  Toennies, J. P. 
Surface Lattice Dynamics and Electron-Phonon Interaction in Cesium Ultra-Thin Films. 
{\em Phys. Chem. Chem. Phys.} {\bf 2017}, 19, 16358-16364.


\bibitem{kF-layer} Calculated from $k_F = \sqrt{3 \pi / a_c n_{sat}}$.


\bibitem{AllenT} Poole, C. P.; Zasadinsky, J. F.; Zasadinsky, R. K.; Allen, P. B. {\em Electron-Phonon Coupling Constants}, in {\em Handbook of Superconductivity}, Poole, C. P., Jr. ed.; Academic Press: New York, 1999. Ch. 9, Sec. G, pp. 478-483.

\bibitem{Grimvall} Grimvall, G. {\em The Electron-Phonon Interaction in Metals}; North-Holland: New York;1981.

\bibitem{k} Michaelson, H. B. Work Function of Elements and its Periodicity. 
{\em  Journal of Applied Physics} {\bf 1977}, 48, 4729-4733.


\bibitem{81Kie} Kiejna, A.; Wojciechowski, K. F. Work Function of Metals: Relation between Theory and Experiment. {\em Prog. Surf. Sci.} {\bf 1981}, 11, 293-338.

\bibitem{Qi} Qi, Y.; Ma, X.; Jiang, P.; Ji, S.; Fu, Y.; Jia, J. F.; Xue, Q.-K.; Zhang, S. B. 
Atomic-Layer-Resolved Local Work Functions of Pb Thin Films and Their
Dependence on Quantum Well States. {\em App. Phys. Lett.} {\bf 2007}, 90, 013109.



\bibitem{Layer2} Flach, B. Thesis; University of G\"ottingen: 2000.
Besides the data for Li(110)/W(110) reproduced in Fig. 3(a), this thesis reports other  unpublished data on HAS drift measurements on 5 and 9 ML Li(110)/W(110) measured at 140 K for $k_i$ varying from 5.2 to 9.2 $\AA^{-1}$, from which a value for $\lambda$ of $0.39 \pm 0.11$ is obtained.


\bibitem{Layer4} Carlsson, A.; Hellsing, B.; Lindgren, S. A.; Walld\'en, L. High-Resolution Photoemission from a Tunable Quantum Well: Cu(111)/Na.  {\em Phys. Rev. B} {\bf 1997}, 56, 
1593-1600. Note: this is a calculated value of $\lambda$ for the surface.

\bibitem{l} Hinch, B. J.; Koziol, C.; Toennies, J.P.; Zhang, G. Evidence for Quantum Size Effects Observed by Helium Atom Scattering During the Growth of Pb on Cu(111).
{\em  Europhys. Lett.} {\bf 1989}, 10, 341-346.


\bibitem{Layer8} Flach, B.; Hulpke, E.; Sternh\"ogle, X . Characterization of Epitaxial Rubidium Films with Helium-Atom Scattering. {\em Surf. Sci.} {\bf 1998}, 412, 12-23.


\bibitem{Beckmann} Beckmann, P.;  Spizzichino, A. {\em The Scattering of Electromagnetic Waves from Rough Surfaces}; The Macmillan Company: New York; 1963, Ch. 5, p. 70.

\bibitem{Layer5} Schief, H.; Toennies, J. P.  Observation of Valence-Band Structure in the LVV-Auger Spectra of Thin Epitaxial Sodium Layers.  {\em Phys. Rev. B.} {\bf 1994}, 50, 8773-8780.


\bibitem{Braun} Braun, J. Thesis; University of G\"ottingen: 1997. Max-Planck-Institut f\"ur Str\"omungsforschung Report.  

\bibitem{GZhang} Zhang, G. Thesis; University of G\"ottingen: 1991. Max-Planck-Institut f\"ur Str\"omungsforschung Report.
    
\bibitem{Chulkov} Sklyadneva, I. Yu.; Heid, R.; Bohnen, K.-P.; Echenique, P. M.;
Chulkov, E. V. Surface Phonons on Pb(111). {\em  J. Phys.: Condens. Matter} {\bf 2012}, 24, 
104004.

\bibitem{Yndurain}  Calleja, F.; V\'azquez de Parga, A. L.; 
Anglada, E.; Hinarejos, J. J.; Miranda,R.; Yndurain, F. Crystallographic and Electronic Contribution 
to the Apparent Step Height in Nanometer-Thin Pb(111) Films Grown on Cu(111). 
{\em New Journal of Physics} {\bf 2009}, 11, 123003.

\bibitem{Layer1} Lang, N. D.;  Kohn, W. Theory of Metal Surfaces - Work Function. 
{\em  Phys. Rev. B} {\bf 1971}, 3, 1215-1223, and references therein.


\bibitem{Zhang-05} Zhang, Y. F.;  Jia, J.-F.; Han, T.-Z.; Tang, Z.;  Shen, Q.-T.; Guo, Y.;  Qiu,
Z. Q.;  Xue, Q.-K. Band Structure and Oscillatory Electron-Phonon Coupling of Pb Thin Films Determined by Atomic-Layer-Resolved Quantum-Well States.  {\em Phys. Rev. Lett.}  {\bf 2005},
95, 096802.

\bibitem{13Skl}  Sklyadneva, I. Yu.; Heid, R.; Bohnen, K.-P.; Echenique, P.M.; Chulkov, E. V.  
Mass Enhancement Parameter in Free-Standing Ultrathin Pb(111) Films: The Effect of Spin-Orbit Coupling. {\em Phys. Rev. B} {\bf 2013}, 87, 085440. 

\bibitem{plummer} Plummer, E. W.; Shi, J.; Tang, S.-J.; Rothenberg, E.; Kevan, S. D. 
Enhanced Electron–Phonon Coupling at Metal Surfaces. {\em  Surf. Sci. Rep.}
{\bf  2003}, 74, 251-268


\bibitem{rusina} Rusina, G. G.;  Borisova, S. D.; Eremeev, S. V.;  Sklyadneva, I. Yu.; 
Chulkov, E. V.; Benedek, G.; Toennies, J.P.  Surface Dynamics of the Wetting Layers and Ultrathin
Films on a Dynamic Substrate: (0.5-4) ML Pb/Cu(111).  {\em J. Phys. Chem. C} {\bf  2016}, 
120, 22304 –22317

\end{thebibliography}







\end{document}